\documentclass[prb,twocolumn,superscriptaddress,preprintnumbers,amsmath,amssymb,floatfix]{revtex4}
\usepackage{graphicx}

\begin{document} 
\title{Electronic correlations in short period (CrAs)$_n$/(GaAs)$_n$ 
ferromagnetic heterostructures}
\author{L. Chioncel}
\affiliation{Augsburg Center for Innovative Technologies, Augsburg, D-86135, Germany}
\affiliation{Institute of Physics, University of Augsburg, D-86135, Germany}
\author{I. Leonov}
\affiliation{Theoretical Physics III, Center for Electronic Correlations and Magnetism, 
Institute of Physics, University of Augsburg, Augsburg 86135, Germany}
\author{H. Allmaier}
\affiliation{Institute of Theoretical and Computational Physics, Graz University of Technology,
A-8010 Graz, Austria}
\author{F. Beiuseanu}
\affiliation{Faculty of Science, University of Oradea, RO-47800, Romania}
\author{E. Arrigoni}
\affiliation{Institute of Theoretical and Computational Physics, Graz University of Technology,
A-8010 Graz, Austria}
\author{T.~Jurcut}
\affiliation{Faculty of Science, University of Oradea, RO-47800, Romania}
\author{W. P\"otz}
\affiliation{Institute of Theoretical Physics, Karl-Franzens University Graz,
A-8010 Graz, Austria}
 
\begin{abstract}
We investigate 
half-metallicity 
in  [001] stacked (CrAs)$_n$/(GaAs)$_n$ heterostructures
with $n \leq 3$ by means of a combined many-body and electronic structure calculation.  
Interface states in the presence of strong electronic correlations are discussed for the
case  $n=1$. For $n=2,3$ our results indicate that the minority spin half-metallic gap
is suppressed by local correlations 
at finite temperatures, and  continuously shrinks upon increasing the heterostructure
period. Although around  room temperature the magnetization
of the heterostructure deviates by only $2\%$ from the 
ideal integer value, finite temperature polarization at $E_F$ is
reduced by at least $25\%$. Below the Fermi
level the minority spin highest valence states are found 
to localize more on the GaAs layers while lowest conduction states have a many-body origin. 
Our results, therefore, suggest that in these heterostructures
holes and electrons remain separated among different layers.
\end{abstract} 

\maketitle 

\section{Introduction}
In recent years, the increasing ability to control the growth of semiconductor crystals 
has made possible the fabrication of high-quality artificial heterostructures of many different
geometries and semiconductor classes. Heterostructures interfacing  half-metals with semiconductors
are technologically very attractive, since in principle they can be used to attain  high polarization
spin injection from a ferromagnetic electrode into the semiconductor. The high polarization  
results from the  main property of half-metallic ferromagnets, namely, the fact that they exhibit a
metallic density of states for one spin channel and a gap at the Fermi-level for the 
other~\cite{gr.mu.83,ka.ir.08}. The major players in the present semiconductor-based 
electronic technology are zinc-blende structures.
For this reason,  half-metals which can adapt to the
structure and bonding of zinc-blende semiconductors are especially attractive as they 
are compatible with existing technology.

Recently, {\it Zhao} and {\it Zunger}~\cite{zh.zu.05} investigated the relative stability of NiAs
and zinc-blende (ZB) structures under pseudomorphic epitaxial conditions.
They found that under epitaxial growth condition, most of
the Cr and Mn pnictides and chalcogenides cannot be stabilized
below a lattice constant of $6.5\AA$. However, these conclusions
are valid for the growth of thick layers, whereas the
growth of very thin films is dominated by the strain
energy at the interface. This explains the experimental observation that CrAs 
can be grown in the zinc-blende structure for thicknesses up to a few monolayers. 

The first experimental realization of such thin films was done by 
{\it Akinaga et al.}~\cite{ak.ma.00}. In their work they have synthesized
zinc-blende CrAs thin films of $3nm$ thickness on GaAs substrates. They measured a magnetic moment of 
$3 \mu_B$, which is in agreement with theoretical prediction, and found an experimental Curie temperature 
above 400 K~\cite{ak.ma.00}. From the experimental point of view, such small thickness 
makes this material difficult to use in practical devices.
Therefore, attention was 
directed towards CrAs/GaAs multilayers~\cite{mi.ak.02}. Previous ab-initio calculations 
showed high spin polarization through the entire region of the multilayer in the case of 
two monolayer CrAs and two monolayer of GaAs stacked alternatively (CrAs)$_2$/(GaAs)$_2$~\cite{na.sh.04}. 
Initially, the produced multilayers of zinc-blende CrAs/GaAs grown on 
GaAs substrates~\cite{mi.ak.02} indicated that the surface and interface of the 
multilayer were not completely flat. However, it was found that the multilayers grow much thicker 
than pure zinc-blende CrAs. Recently, by optimizing  the growth temperature, the quality 
was improved significantly, and epitaxial growth of zinc-blende multilayer with flat surface 
and interface was achieved~\cite{ak.mi.04}. The magnetization measurements for a multilayer
structure of (CrAs)$_2$/(GaAs)$_2$ repeated 100 times~\cite{ak.mi.04} showed a value of 
$2\mu_B$ per formula unit, lower than the theoretical prediction of $3\mu_B$. In addition,
the temperature dependence of magnetization indicated a ferromagnetic transition temperature
of about 800K.  
Moreover, it was confirmed by electronic structure calculations 
that the spin polarization is preserved throughout the multilayer, and 
that it is insensitive to substitutional disorder between the Cr and Ga sites~\cite{na.sh.04}. 

It is the purpose of the present paper to investigate effects caused by many-body correlations
at finite temperatures in the short period (CrAs)$_n$/(GaAs)$_n$ heterostructures.
These ab-initio calculations are performed within a combined density functional and
many-body dynamical mean field theory.
Our results show that correlations do not affect much 
magnetisation in these materials, while polarisation is strongly suppressed.
In particular,  the minority spin gap contributes in confining the electrons and
holes in the heterostructure,  an effect similar to the one discussed for the
(GaAs)$_n$/(AlAs)$_n$superlatices where  electrons and  holes
are spatially separated~\cite{go.ch.89}.

The paper is organized as follows: the microscopic description of 
electronic states around the half-metallic gap in  bulk CrAs  
is briefly summarized in  Section \ref{band-edge-bulk}.
The geometry of multilayers structures which results in lowering the symmetry 
are discussed in Section \ref{multi-sym}. Results are presented in
Section \ref{results}. 

\section{Band-edge states in half-metallic bulk CrAs} 
\label{band-edge-bulk}
Before discussing the nature of the half-metallic gap in (CrAs)$_n$/(GaAs)$_n$ heterostructure 
with or without  
electron-electron interactions, let us briefly summarize the principal physical factors 
leading to gap formation in the zinc-blende bulk materials. 
\begin{figure}[h]
\includegraphics[width=1.00\columnwidth,angle=-90]{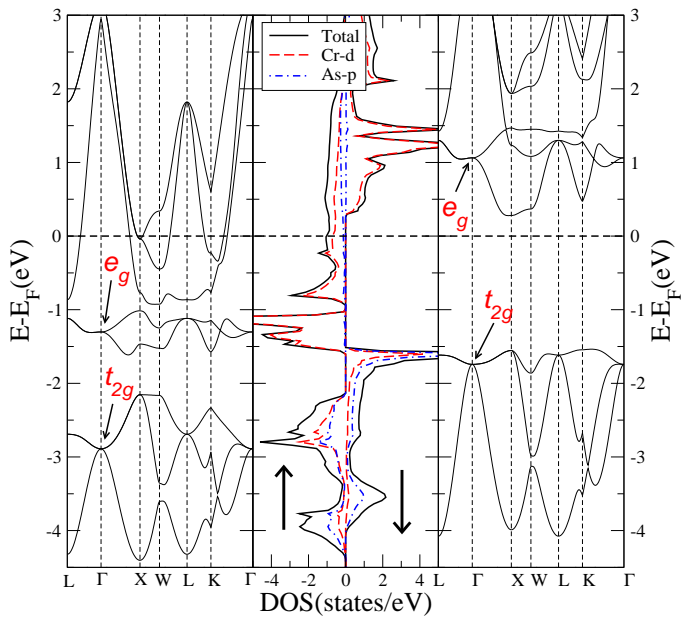}
\caption{(Color online) Local-Spin Density Approximation (LSDA) 
 - total (black line) and atom resolved (Cr(dashed red) As(doted dashed blue)) 
density of states for bulk CrAs with a lattice parameter 
a=5.75$\AA$ (central panel). Left and right panels display the majority and 
minority spin band structure, respectively. States with $t_{2g}$ and $e_g$
symmetry are indicated by an arrow at the $\Gamma$ point.}
\end{figure}

\begin{figure}[h]
\includegraphics[width=1.00\columnwidth]{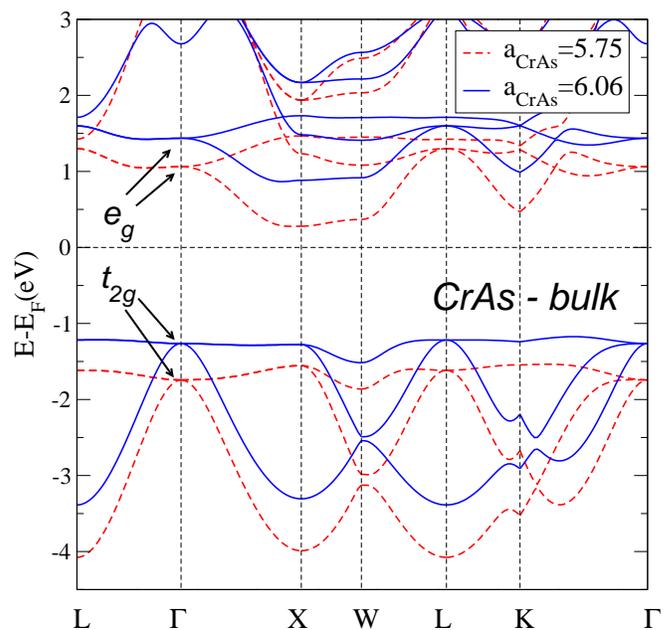}
\caption{(Color online) Minority spin bulk-CrAs band structure 
for the different lattice parameters: $a=5.75 \AA$ corresponding to an
enlarged CrAs (long dashed-red), and $a=6.06 \AA$ corresponding to 
the InAs lattice parameter (solid-blue). By increasing the lattice
parameter the gap changes from a  direct to an indirect one.
Results obtained from a LSDA calculation.
\label{gaas_cras_bulk}}
\end{figure}

Bulk properties of zinc-blende pnictides and chalcogenides were discussed in many 
papers~\cite{co.pi.01,zh.ge.02,xi.xu.03,ga.ma.03}. 
In particular, first-principles calculations \cite{shir.03,ak.ma.00} predicted CrAs to be half metallic.
In this structure every atom has a tetrahedral 
coordination with the first neighbors being of the other atomic species.  $d-$states
split into the $t_{2g}$ and $e_g$ manifolds. While  $t_{2g}$-states hybridize with the
$p$-states of the neighboring atom, forming bonding and anti-bonding states, 
 $e_g$ orbitals are practically non-bonding and form narrow bands. The 
bonding / anti-bonding splitting is a characteristic 
of the tetrahedral coordination. The Fermi-level E$_F$ is situated in the gap between the bonding bands and the narrow non-bonding 
e$_g$-bands. In addition, the existence of the band gap is assisted by the exchange splitting 
which keeps minority e$_g$ states higher in energy. 
A preliminary analysis of the electronic bands in the minority spin channel 
shows that the nature of the gap can change depending on the 
lattice parameter. For a=5.75$\AA$, Fig. \ref{gaas_cras_bulk} shows a ``direct gap'' formed at the X-point,
while for a=6.06$\AA$ the gap appears to have an ``indirect'' nature. 
In addition, we verified that
states at the top of the valence band are p-d hybridized states, 
while states at the bottom of the conduction band (electrons) are predominantly Cr-d states.
As one can see in Fig. \ref{gaas_cras_bulk} the entire manifold around
the Fermi level is shifted to higher energies as the lattice parameter is increased.
Such changes in the lattice parameter may occur 
when the half-metal is grown on a semiconductor 
substrate. 

Electronic structure calculations for bulk CrAs for different lattice constants at finite 
temperatures and including correlation effects were performed recently
by one of the authors~\cite{ch.ka.05}. 
It was shown that correlations induce spectral
weight in the minority spin gap, while the material remains half-metallic.
The spectral weight in the minority spin gap is known as so-called non-quasiparticle (NQP) states.
The occurrence of these states is connected to ``spin-polaron''
processes: the spin-down low-energy electron excitations,
which are forbidden for half-metallic ferromagnets in the one-particle picture, turn
out to be possible as superpositions of spin-up electron excitations and virtual
magnons~\cite{ed.he.73,ir.ka.90}. It is important to mention that the only situation 
in which 
half-metallicity 
is preserved in the presence of finite temperature correlations
is in CrAs and VAs, while in  other (semi-)Heusler materials 
the additional spectral weight builds up at the Fermi energy, so that 
strong depolarization takes place~\cite{ch.ka.03,ch.al.07,ch.sa.08}. 
Depolarisation can  originate from other effects, such as 
spin-orbit coupling, which causes 
a mixing of the two spin channels. However, in  CrAs this effect was
found to be less than 
$1\%$~\cite{sh.ik.03}.
Consequently, this interaction is not taken into account for 
the present calculations. 

There is enormous literature on GaAs band-structure calculations,
all results showing a direct gap at the $\Gamma$ point in the band structure. 
The general discussion 
is focused on the difficulty of 
density-functional theory (DFT) 
mean-field type calculations based on 
Local-Density Approximation (LDA)/
Generalised-Gradient Approach (GGA)
to reproduce the magnitude of the experimental gap which is around $1eV$. 
It has been assumed that the band gap problem does not occur in half-metals since
their dielectric response is that of a metal. This assumption was
recently contradicted by a  GW -type of calculation for 
La$_{0.7}$Sr$_{0.3}$MnO$_3$, 
which predicts a half-metallic band gap that is 2 eV larger 
than the one obtained by DFT\cite{ki.ar.03}.
In particular, for the present 
case of semiconducting and half-metal heterostructures this might imply that the
DFT gaps might  underestimate the actual experimental values.

In the present work we investigate the nature of the minority spin states involved
in the gap formation of (CrAs)$_n$/(GaAs)$_n$ heterostructures. In our analysis
the CrAs bulk states situated at the X-point play a crucial role because by
Brillouin zone folding  associated with symmetry lowering present in the 
(CrAs)$_n$/(GaAs)$_n$ heterostructure, the X point is 
folded into the $\Gamma$ point, where the bottom of the conduction band in GaAs is 
expected to be present. Therefore, the relative position of the states in X-point 
with respect to the bottom of the GaAs conduction band, would contribute 
to the band-edge states. This effect will be discussed in detail in the next sections.
Moreover, we investigate up to what extent
 the electronic states forming the (CrAs)$_n$/(GaAs)$_n$ 
band structure are significantly changed by finite temperatures dynamic correlations.

\section{Multilayers geometry and  method of calculation} 
\label{multi-sym}
\begin{figure}[h]
\includegraphics[width=0.99\columnwidth]{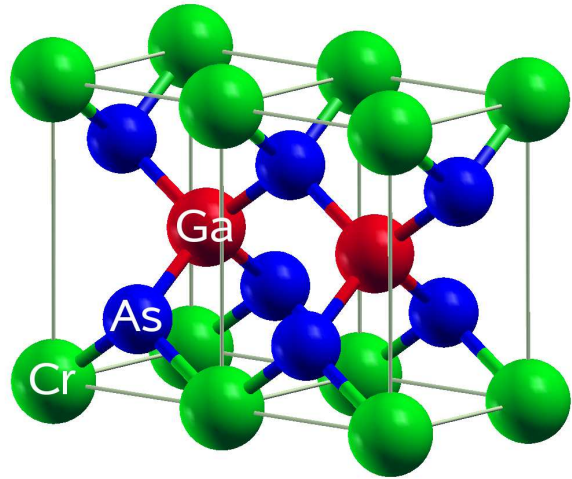} 
\caption{(Color online) $(CrAs)_1/(GaAs)_1$ supercell, which is repeated twice along the $x$-axis
for illustrative purposes. Atoms situated in different symmetry positions are presented 
by colored spheres Cr (green) - lower and upper planes, As (blue) - above and below the Cr layers 
and Ga (red) -  center  of the unit cell. }
\label{scell_11}
\end{figure}

The (CrAs)$_n$/(GaAs)$_n$ heterostructure, with $n \le 3$ 
extends  along the $[001]$ direction of the of bulk zinc-blende material.
The unit cells of the superlattice having the space group symmetry $D_{2d}^1$ 
are simple tetragonal with $c/a_0=2n/\sqrt{2}$, where c is the unit-cell dimension 
along the stacking direction. The basis of the tetragonal cell have the 
constants $a=b=a_0/\sqrt2$, where $a_0$ represents the FCC lattice parameter. 
The positions of the atoms in the (CrAs)$_1$/(GaAs)$_1$ unit supercell
are  Cr:$a/2(0,0,0)$, As:$a/2(0,1,c/a)$, Ga:$a/2(1,1,c/a)$ and 
As:$a/2(1,0,3c/2a)$ (see Fig. \ref{scell_11}). In the heterostructure
calculations  we included
empty spheres with no net nuclear charge in the empty tetrahedral sites in order to obtain
a close-packed structure,
as it is generically done for zinc-blende semiconductors.
For the (CrAs)$_1$/(GaAs)$_1$ case the 
empty spheres surrounding the cations and anions are located 
at $a/2(1,0,3c/2a)$, $a/2(0,1,c/2a)$, $a/2(0,0,c/a)$, $a/2(1,1,0)$. The positions
of the atoms in larger unit cells for the other superlattices can be written by
extrapolating the (CrAs)$_1$/(GaAs)$_1$ case.

Similar supercells were considered previously in 
discussing the electronic structure of semiconducting (GaAs)$_m$/(AlAs)$_n$ 
superlattices~\cite{go.ch.89}. For the particular case when $m=n$ it was demonstrated 
that within the tetragonal Brillouin zone the X point  (along the {\bf k}$_z$ direction)
of the FCC zone folds onto the $\Gamma$ point. 
Upon increasing the number of layers in
the superlattice, the Brillouin zone is compressed along {\bf k}$_z$ and the number
of zone foldings increase. Contrary to the bulk semiconductors where the top 
of the valence band is mainly As-p like states, in the case of (GaAs)$_m$/(AlAs)$_n$ 
heterostructures As-p anionic bonds are shared between Ga and Al atoms, leading to 
confinement effects at the valence band maxima. 

For the case of (CrAs)$_n$/(GaAs)$_n$ heterostructures the As-p states are 
shared by Ga and Cr atoms. 
In particular, due to hybridisation the Cr-d states contribute significantly 
to the top of the minority spin valence band. Therefore, in addition 
to the confining effect determined by the dimensionality of the heterostructure, 
electronic correlation are expected to influence the minority spin band-edge states
around the gap.  

In our calculations
we used the lattice parameter $a_0=5.75\AA$, which is the optimized GaAs lattice constant 
obtained from a spin-GGA calculation~\cite{ge.sh.06}. This value is slightly larger than 
the one found experimentally for GaAs 
($a_0=5.65\AA$) and the one predicted for (CrAs)$_2$/(GaAs)$_2$  double monolayers ($a_0=5.69\AA$)~\cite{na.sh.04}. 
We checked that our results do not change when 
the atomic sphere radii 
are chosen to be equal to the average Wigner Seitz radius $R=2.675 a.u.$, in comparison to 
the case were the atomic radii are changed 
by $\pm 5\%$ depending on their type.

To investigate the effect of electronic correlations for the above supercells,
we performed calculations using a recently developed LSDA+
Dynamical Mean Field (DMFT)
 scheme~\cite{ch.vi.03}. 
Correlation effects in the valence Cr-$d$ orbitals are 
included via an on-site electron-electron interaction in the form
$\frac{1}{2}\sum_{{i \{m, \sigma \} }} U_{mm'm''m'''}
 c^{\dag}_{im\sigma}c^{\dag}_{im'\sigma'}c_{im'''\sigma'}c_{im''\sigma} $.
The interaction is treated in the framework of
DMFT
\cite{ko.vo.04,ko.sa.06,held.07}, with a spin-polarized T-matrix Fluctuation Exchange (SPTF) type
of impurity solver \cite{ka.li.02}. Here, $c_{im\sigma}/c^\dagger_{im\sigma}$
destroys/creates an electron with spin $\sigma$ on orbital $m$ on site $i$.
The Coulomb matrix elements $U_{mm'm''m'''}$ 
can be computed for the particular 
material by taking into account the symmetry of the orbitals and the crystal structure
in terms of effective Slater integrals and Racah or Kanamori coefficients~\cite{im.fu.98,ko.sa.06}.
We used the following effective Slater parameters: $F^0=2$eV, $F^2=5.17$eV and $F^4=3.233$eV 
(which results in a Coulomb-interaction of $U=2$eV and a Hund's rule coupling of 
$J=0.6$eV); these are in agreement with previous works~\cite{ch.ka.03,ch.ka.05,ch.ar.06,ch.ma.06}.
%
For the exchange correlation functional the LSDA-approximation 
was used, as we found no significant differences with respect to results using GGA.

Since static correlations are already included in the local spin-density
approximation (LSDA), ``double counted'' terms must be subtracted. To obtain this,
we replace $\Sigma_{\sigma}(E)$ with $\Sigma_{\sigma}(E)-\Sigma_{\sigma}(0)$~\cite{li.ka.01}
in all equations of the LSDA+DMFT procedure~\cite{ko.sa.06}.
Physically, this is related to the fact that DMFT only adds {\it dynamical}
correlations to the LSDA result.
For this reason, it is believed that this kind
of double-counting subtraction ``$\Sigma(0)$'' is more appropriate for a DMFT
treatment of metals than the alternative static ``Hartree-Fock'' (HF) subtraction~\cite{pe.ma.03}.

In the calculations we used 287 (6/4) $k$-vectors for the Brillouin-zone 
integration and used a cutoff of l$_{max}$=6 for the multipole expansion in charge 
density and the potential as well as a cutoff of l$_{max}$=4 for the wavefunctions. 
We checked higher cutoffs for $(CrAs)_1/(GaAs)_1$ and found only negligible 
differences. 

\section{Results} 
\label{results}
\subsection{(CrAs)$_1$/(GaAs)$_1$, heterostructure}
\begin{figure}[h]
\includegraphics[width=0.99\columnwidth]{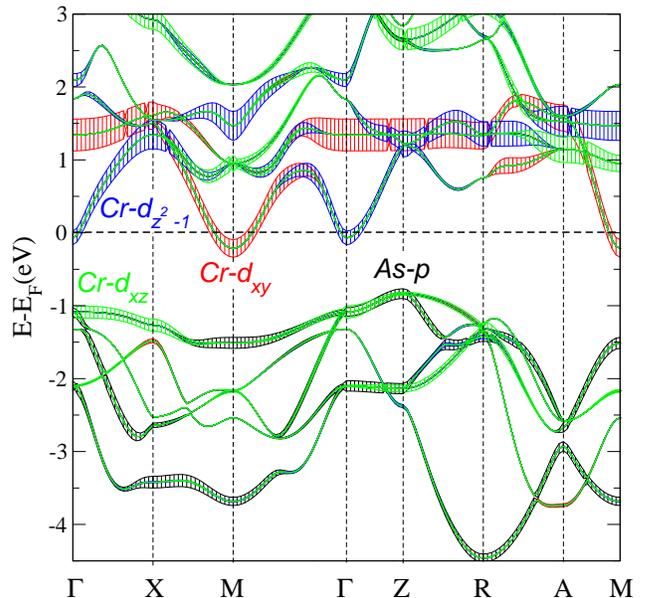}
\caption{(Color online) LSDA minority spin bands for $(CrAs)_1/(GaAs)_1$ along the symmetry-points
in the tetragonal Brillouin zone. 
Bands are decorated according to dominant orbitals character. The top of the 
valence band has mainly an As-p character (black), while the bottom of the conduction band
has Cr-d$_{xy}$ (red) character. Cr-d$_{xy}$  and d$_{z^2-1}$ (blue) orbitals 
cross the Fermi level. 
\label{bnds_1-1}}
\end{figure}

The calculated LSDA-band structure of (CrAs)$_1$/(GaAs)$_1$ displays an overall metallic
behavior. It is of special interest to investigate the minority spin channel band 
structure shown in Fig. \ref{bnds_1-1}. We plot results in the energy range of 
$-5$ to $3eV$ in order to distinguish between the orbitals
around the Fermi level. In the tetragonal Brillouin zone the symmetry points 
are: $\Gamma=(\pi/a)(0,0,0)$, $R=(\pi/a)(1,0,a/2)$, $A=(\pi/a)(1,1,a/c)$,
$Z=(\pi/a)(0,0,a/c)$, $M=(\pi/a)(1,1,0)$ and $X=(\pi/a)(1,0,0)$.
At lower energies (not shown)  between $-12$ and $-10eV$ one can observe a
s-like lower valence band of 2eV width. This 
is separated from the upper valence band by an interband of 3.7eV. As one can 
see in Fig. \ref{bnds_1-1}  the bands in  the energy range of $-5, -1eV$ are
mainly hybridized As-p and Cr-d orbitals.
At energies around $E_F-1eV$
the As-p states are hybridized with Cr-d$_{xz}$ and are situated at higher
energies at the $Z$-point, with a higher As-p weight. In comparison at the
top of the valence band in the $\Gamma$ point a larger weight is visible for Cr-d$_{xz}$ states.
Here, the As-p contribution is significantly reduced.

States around and just above the Fermi level, also show interesting features. 
As we discussed above due to symmetry reduction, the X-point along {\bf k}$_z$
of the bulk FCC zone folds over to the $\Gamma$ point of the heterostructure zone.
Since the lowest conduction band for the minority spin spin of bulk CrAs is situated 
at lower energy than the conductoin band edge of bulk GaAs at the $\Gamma$ point,
one would expect that in the (CrAs)$_1$/(GaAs)$_1$ heterostructure
the lowest states just above the Fermi level  are an X-like CrAs band. It is 
clearly seen in Fig. \ref{bnds_1-1}, that this is not the case, the Fermi level 
being crossed by Cr-d$_{xy}$ as well as d$_{z^2-1}$ orbitals. 
Note that, for precisely this lattice parameter the bulk CrAs and equivalently 
the (CrAs)$_1$/(CrAs)$_1$ is half-metallic. Therefore the presence of GaAs
determines that Cr-d$_{xy}$ and d$_{z^2-1}$ orbitals 
from the CrAs -- which constitutes the ``interface-layer'' -- to cross the Fermi level.
We have performed selfconsistent total energy calculations
for a fixed volume of the unit cell (lattice parameter $a=5.75\AA$)
as a function of bond length, more  precisely as a function of
the distance between Cr-As and Ga-As atoms
along the stacking direction $\delta=z_{As}-z_{cr}$ which provides information 
concerning the stability of the GaAs layers with respect to the CrAs ones. 
\begin{figure}[h]
\includegraphics[width=0.99\columnwidth]{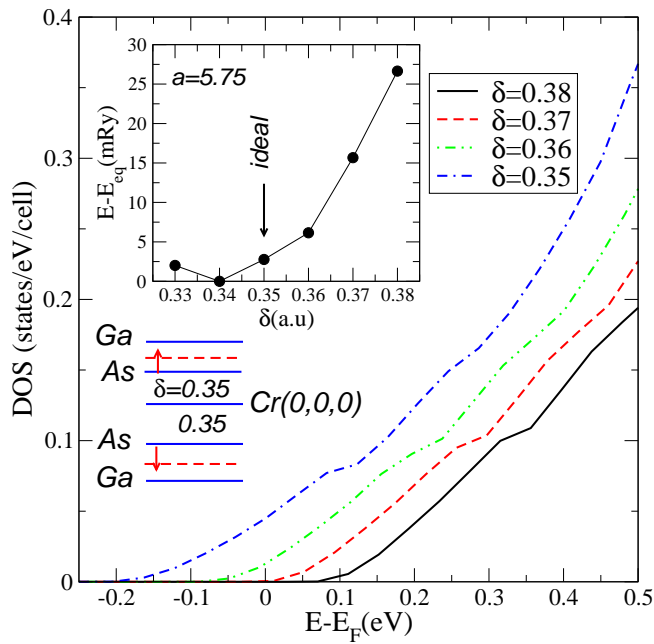}
\caption{(Color online) Evolution of LSDA - Cr-d$_{xy}$ minority spin density of states 
of $(CrAs)_1/(GaAs)_1$ for different distances ($\delta=z_{As}-z_{cr}$)
between As and the Cr layer at fixed volume of the unit cell. The inset shows
a plot of the energy (measured with respect to the equilibrium energy $E_{eq}$)
versus $\delta$. As one can see, at the optimal $\delta=0.34$ the
minority spin gap is closed and the heterostructure is metallic.
\label{gap_dist}}
\end{figure}
In addition this analysis provides also the evolution of the Cr-d$_{xy}$ states within 
the minority spin gap. As one can see in Fig. \ref{gap_dist} within the LSDA a larger 
Cr-As bond length (increasing $z_{As}$) and smaller GaAs bond favor 
half-metallicity.
The results of the total energy calculations show that the equilibrium bond length 
is realized for the distance of $\delta \approx 0.34$ for which a metallic 
(CrAs)$_1$/(CrAs)$_1$ heterostructure is obtained. We have checked that
LSDA+U calculations with  U=2eV and J=0.6eV provide the similar value
for the equilibrium distance. 

\begin{figure}[h]
\includegraphics[width=0.99\columnwidth]{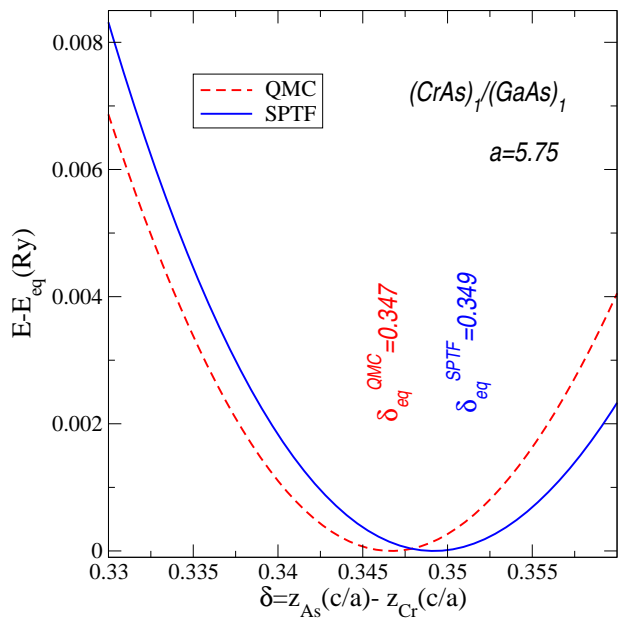}
\caption{(Color online) Total energies for different displacements of GaAs layer
with respect the the CrAs layer in the $n=1$ heterostructure
computed within SPTF (blue solid line) and QMC (red dashed line)
solvers of the DMFT, for U=2eV, J=0.6eV.
\label{etot_dmft}}
\end{figure}

Similar calculations were performed including 
correlation effects captured by DMFT using two different many-body solvers. 
In fig. \ref{etot_dmft} we show the comparison between the calculations 
performed using the SPTF and the QMC -- Hirsch-Fye \cite{jarr.92} solvers. The contribution to the total energy
coming from correlations was computed within the LSDA+DMFT-SPTF as an additional
Galitskii-Migdal-type contribution $\frac 1 2 Tr \hat{\Sigma} \hat{G}$~\cite{ch.vi.03,ma.mi.09}, while the interaction 
energy within the LSDA+DMFT-QMC is computed from the double occupancy matrix 
$\langle \hat{n}_{im\sigma} \hat{n}_{im'\sigma'} \rangle$\cite{he.mc.01,ch.bu.06,le.bi.08,le.ko.10}.  
As one can see in Fig. \ref{etot_dmft} the two numerical results are in good agreement.
Including correlation effects the equilibrium distance between the GaAs and the 
CrAs planes approaches the ideal value of $\delta=0.35$
(corresponding to the FCC structure), thus distortions are 
not favored in this case. Therefore, in what follows we shall consider the 
ideal structure with equidistant GaAs and CrAs planes within the supercells.
Because of the large supercell the following many-body results were obtained 
using the SPTF solver which implements the fully rotational invariant interaction discussed
in the previous section.

\begin{figure}[h]
\includegraphics[width=0.99\columnwidth]{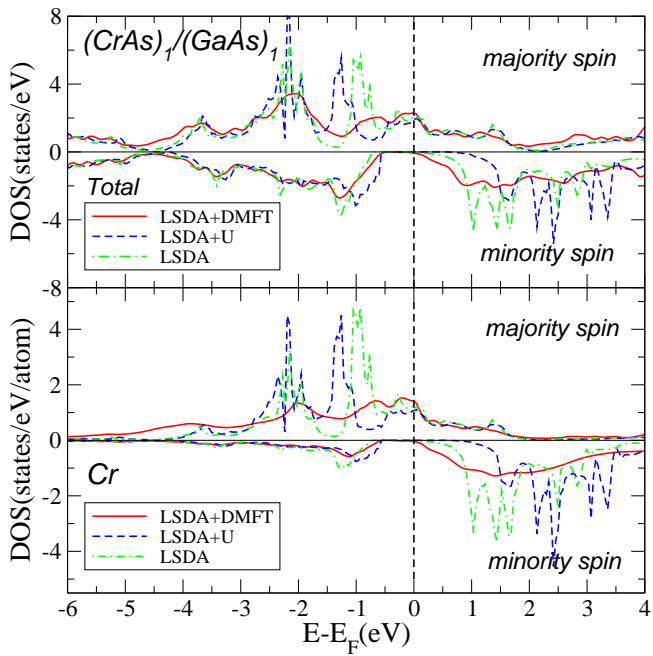}
\caption{(Color online) Total and Cr density of states obtained by LSDA(green dashed-doted line), 
LSDA+U (blue dashed line) and LDA+DMFT(red solid line) for (CrAs)$_1$/(GaAs)$_1$. \label{dos_1-1}
Results obtained for the ideal $\delta=0.35$.}
\end{figure}

In Fig. \ref{dos_1-1} we
compare non-correlated (LSDA),  mean field local but static correlations
(LSDA+U), with dynamic (LSDA+DMFT) results. 
All calculations accounting for correlations were performed for 
the same values of the Coulomb parameter U=2eV and J=0.6eV.  
The minority spin gap 
opens within the LSDA+U calculation,  in contrast to the LSDA+DMFT
results where spectral weight is present in the minority spin gap. 
As one can see in Fig.\ref{dos_1-1} in the total density of states (upper
panel) the Cr states play the dominant role. 
In the occupied part of the majority spin channel, the LSDA states around 
$E_F-1eV$ are slightly shifted towards lower energies in LSDA+U, while in LSDA+DMFT
the same states are shifted towards the Fermi level, 
although the spectral weight is considerably reduced. 
At the Fermi level, similar values for the density of states are obtained in
both mean-field calculations, while a slightly larger values are obtained including
dynamic correlations. For the total density of states, $E_F \pm 1eV$ is the 
energy range were significant differences can be seen. 
These differences are attributed to Cr density of states shown in
the lower panel of Fig. \ref{dos_1-1}.

\begin{figure}[h]
\includegraphics[width=0.99\columnwidth]{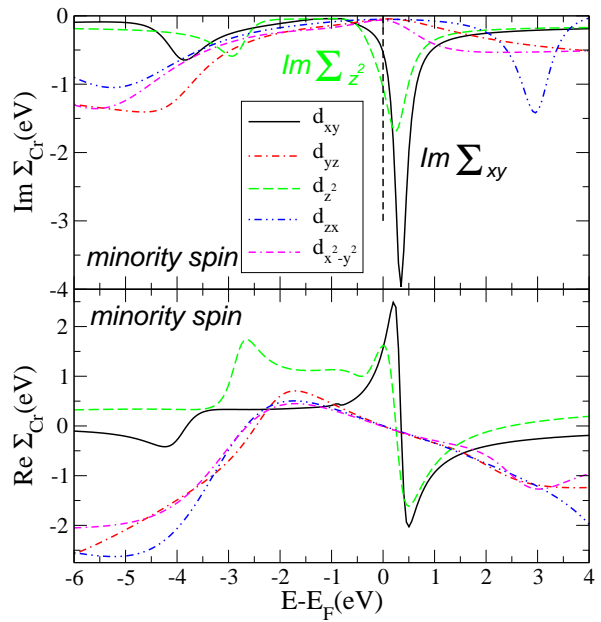}
\caption{(Color online)
Real and imaginary part of the  self energies for Cr orbitals in the $n=1$ heterostructure.
Large values of $\Im \Sigma$ for  Cr-d$_{xy}$ (black solid line) and -d$_{z^2-1}$ (dashed line)
just above the Fermi level are related to the occurrence of NQP states.\label{sig_1-1}}
\end{figure}

The LSDA results give a magnetic moment close to an integer $2.99 \mu_B$ and 
by including U at the mean-field level an integer magnetic moment 
of $3\mu_B$ and a gap of $0.94eV$ is obtained.  
The DMFT density of states is shifted 
towards $E_F$, with states just above the Fermi level. Despite their slight 
spectral weight close to the Fermi level, the magnetic moment has a non integer
value of $2.88\mu_B$. We attribute this reduction of the magnetic moment to the 
appearance of non-quasiparticle states. To discuss further this effect
we plot in Fig. \ref{sig_1-1} the imaginary/real  part
of minority self-energy for all Cr-d orbitals. One can see a 
strong imaginary part associated with a strong energy dependence  of the real
part predominantly in Cr-d$_{xy}$ and  Cr-d$_{z^2}$ orbitals
just above the Fermi level which allows to identify 
the character of non-quasiparticles states.                      
 
The presence of NQP states have been shown previously in the bulk CrAs calculations
\cite{ch.ka.05}. Their position in bulk is situated at higher energies above the Fermi level, 
in comparison with (CrAs)$_1$/(GaAs)$_1$, so that no tails cross the Fermi level 
and an integer magnetic moment of $3\mu_B$ is obtained at finite temperatures.
From the above results we conclude that the presence of GaAs layers triggers the 
presence of ``interface'' Cr-d$_{xy}$ and Cr-d$_{z^2-1}$ states with very small spectral
weight within the minority spin channel at the Fermi level. In the presence of correlations
NQP states are formed on these orbitals.        

\subsection{(CrAs)$_n$/(GaAs)$_n$ heterostructure, with n=2,3}
For $n=2$ the two Cr atoms are equivalent, while the As atoms in this heterostructure
are shared either by two Cr atoms, two Ga atoms or one Cr and one Ga atom. In this 
case the heterostructure can be viewed as two (CrAs)$_1$/(GaAs)$_1$ interfaces coupled
to the GaAs or CrAs end of the layers. Increasing the heterostructure further to $n=3$, 
makes that Cr atoms become inequivalent, having one inner Cr layer and two equivalent 
outer Cr layers. 

For the (CrAs)$_n$/(GaAs)$_n$ heterostructures with $n=2,3$ a self-consistent LSDA calculation
yields a half-metallic solution with an integer magnetic moment of 
$3\mu_B$ per Cr atom. The minority spin gap is $0.94eV$ for $n=2$ and  $0.75eV$ for $n=3$.
The minority spin gap is $0.94eV$ for $n=2$ and  $0.75eV$ for $n=3$.
Within LDA+U, the gap is slightly enlarged to $1.1eV$ ($n=2$), and
$0.93eV$ ($n=3$), while the values for the magnetic moments remain unchanged.
\begin{figure}[h]
\includegraphics[width=0.99\columnwidth]{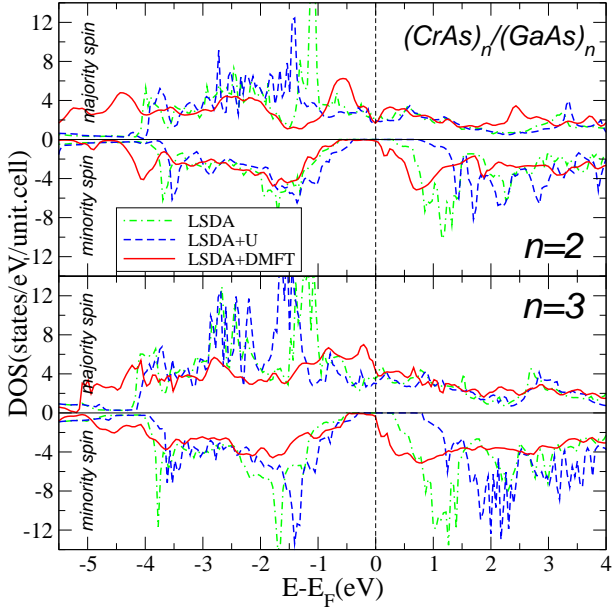}
\caption{(Color online) Total density of states for (CrAs)$_n$/(GaAs)$_n$  
heterostructures for $n=2$ and $3$.  
LSDA (green - dashed doted line) and LSDA+U (blue - dashed line) calculation yield an half-metallic state. 
 DMFT (red -solid line) results suggest the existence of NQP states just 
above the Fermi level with a very small spectral weight at $E_F$. \label{dos_22_33}}
\end{figure}

DMFT results for $n=2$ show in the majority spin channel 
a redistributions of spectral weight: the LSDA/LSDA+U peak at $-1/-1.5eV$
is shifted to lower energies, where new structures appear at higher bonding 
energies around $-5eV$. At the Fermi level all three methods gives a similar value
for the density of states. In the minority spin channel NQP states are visible just above
the Fermi level, while below $E_F$ the band offset is similar to the LSDA 
value. In fig. \ref{dos_22_33} one can compare the overall correlation effects 
captured at different levels. Within  LSDA+U,  the gap is further increased as 
the bandwidth shrinks.
On the other hand, dynamic correlations reduce the gap and enlarge
the band width creating high-bond energy states and evidence the presence of 
correlation induced many-body states in the close vicinity of the Fermi level.
Similar results are obtained for  $n=3$. The computed values of the magnetic moment
are $5.96$ and $8.99\mu_B$ for $n=2$ and $n=3$ respectively at 200K. At 300K
magnetic moments do not change significantly $\mu_{n=2/3}=5.97/8.98$ and the 
spin polarization computed at $E_F$ is again almost temperature
independent and has  the values of $P_{DMFT}^{n=2}\approx 0.76$ and
$P_{DMFT}^{n=3}\approx 0.71$.
 
\begin{figure}[h]
\includegraphics[width=0.99\columnwidth]{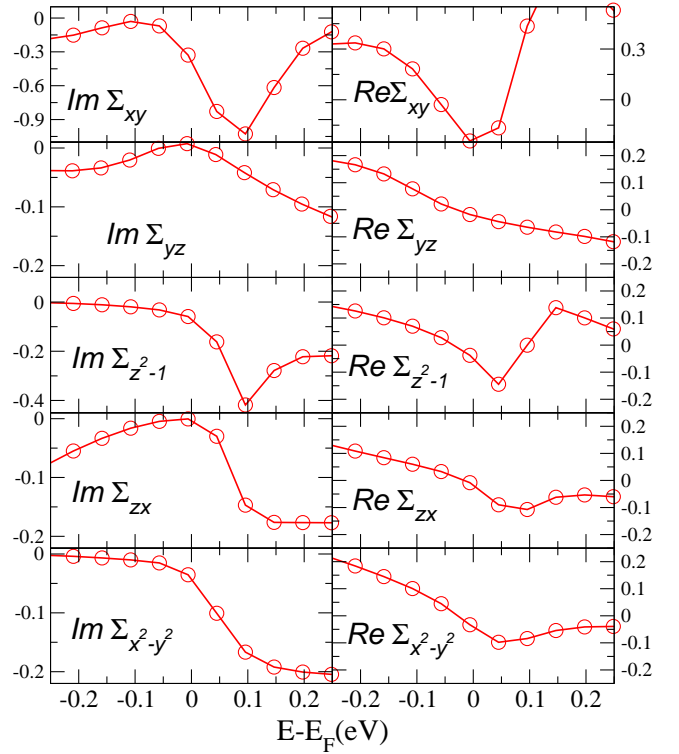} 
\caption{(Color online) Atom and orbital resolved minority spin self-energies 
for (CrAs)$_2$/(GaAs)$_2$ heterostructure for U=2eV, J=0.6eV  and T=200K. 
${\mathit Im} \Sigma$/${\mathit Re} \Sigma$ are plotted on the left/right
column.  
\label{sig_22}}
\end{figure}
As a consequence of geometrical relations between the heterostructures with 
$n=1$ and $n=2$ it is expected that also correlation effects would not be very different. 
In fig.\ref{sig_22} we show the layer- and orbital-resolved 
imaginary part of the minority spin self-energy for the heterostructure with $n=2$.
The self-energy plots contain always a single curve because of the equivalence of 
Cr atoms. In comparing to Fig. \ref{sig_1-1} one can see again that the 
Cr-d$_{xy}$ and Cr-d$_{z^2-1}$ states would present similar departures from 
the usual Fermi liquid description, although in this case 
the amplitude of the self-energy anomaly
above $E_F$ is reduced. This reduction can be attributed 
to the fact that increasing $n$ by adding layers allows for hybridization of 
CrAs-GaAs layers 
which softens 
the self-energy anomaly
present in 
the single correlated magnetic Cr layer in the 
$n=1$ heterostructure. 

\begin{figure}[h]
\includegraphics[width=0.99\columnwidth]{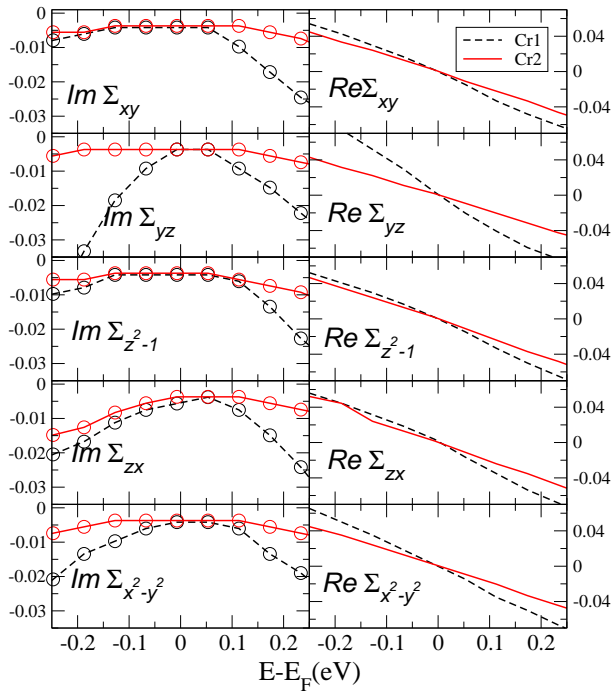} 
\caption{(Color online) Atom and orbital resolved minority spin self-energies 
for (CrAs)$_3$/(GaAs)$_3$ heterostructure for U=2eV, J=0.6eV  and T=200K. 
${\mathit Im} \Sigma$/${\mathit Re} \Sigma$ 
are plotted on the left/right column. Cr1(interface)/Cr2(inner) atoms are plotted in 
dashed-black/red-solid.  An almost linear behavior can be seen
for ${\mathit Re} \Sigma$ for inner type atoms. Possible kinks
are visible for ${\mathit Re} \Sigma_{Cr1}$. 
\label{sig_33}}
\end{figure}
In Fig. \ref{sig_33} we present the minority spin imaginary and real
parts of the Cr1 (inner layer) and Cr2 (interface layer) selfenergies.
To analyze the correlation effects taking place for different $n$, we
discuss a comparison of the energy dependence of the single particle 
self-energies in Figs. \ref{sig_1-1}, \ref{sig_22}, \ref{sig_33}.
The visible differences are meaningful from a physical point of view 
as a results of the increasing number of Cr -layers that can couple with 
increasing $n$. For n=1 the system consists of a
correlated single magnetic layer, for which the 
self-energy apparently deviates
from the Fermi liquid description. For $n=2$ two such Cr correlated layers couple and
the 
self-energy anomalies
are softened. 
Finally,  the result
for $n=3$ does not show 
anomalous
features in the energy dependence of the self-energy.  
Upon looking for small
energies around the Fermi level, one can qualitatively discuss the quasi-particle
lifetime or its inverse the scattering rate $1/\tau_{\bf k} \propto - \Im \Sigma(E_F)$. 
In comparing directly Figs. \ref{sig_22}, \ref{sig_33} the curvature of $\Im \Sigma$
is much larger for $n=2$ 
than
for $n=3$. It is also important to note that for $n=3$
the curvature is considerably larger for the interface layer, than for the inner layer. 
The above qualitative analysis of the energy dependence of the self-energy curves 
suggest clearly that at the interface correlation effects are more 
important leading to NQP sates for the n=1 and n=2 cases, while for n=3 Cr-d states can 
be described by regular quasiparticle states.

\begin{figure}[h]
\includegraphics[width=0.99\columnwidth]{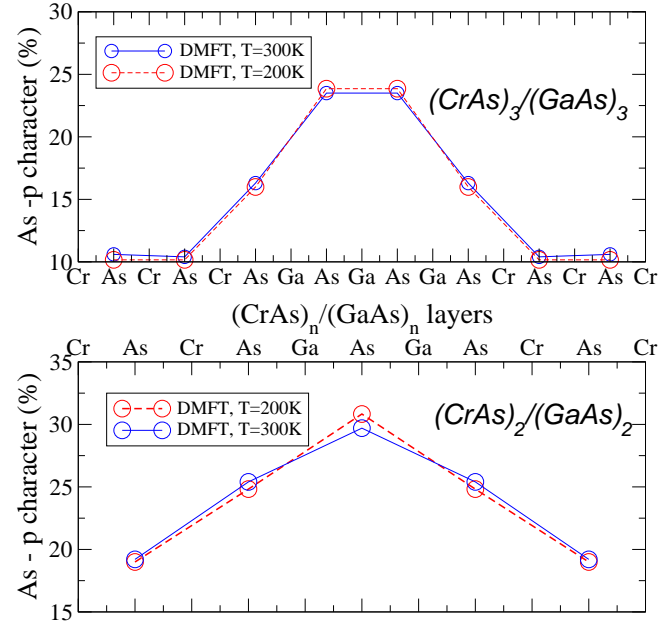}
\caption{(Color online) Relative contribution (in percent) of As-p states 
to the valence band-maxima for each layer of the two superlattices
($n=2$ and $3$). Results are shown for
(CrAs)$_n$/(GaAs)$_n$ heterostructures with U=2eV, J=0.6eV  and T=200K (red-dashed), 
300K (blue-solid).
\label{asp_22-33}}
\end{figure}

In order to study the band edges of the minority spin channel,  we investigate
the character (angular momentum composition) of the electronic states below and above $E_F$. 
As discussed above, the bottom of the conduction band consists of NQP states situated
just above $E_F$, while the top of the valence band is predominantly As-p character. 
We now discuss the
contributions of As-p states originating from different As layers to
the top of minority spin valence band.  Fig. \ref{asp_22-33} shows the 
As-p contributions to the top of the valence band as a function of numbers of layers.
As the distance between the transition metal
mono-layer and As mono-layer increases, the As-$p$
character continues to increase. At the same time, the difference between
values from different monolayers decreases. These results suggest that the
As layer situated closer to the transition metal has its electrons
confined by the $p-d$ hybridization. At larger distances, hybridization
decreases, and more $p$-character is available to form the top of the valence
band. 
At larger temperatures, 
the As-p relative contribution to the top of the valence band slightly 
decreases within the GaAs-layers and increases for As contained within the
CrAs layers. This demonstrates that the confinement of carriers seen 
already at the LSDA level, i.e. 
valence-band top almost confined on the GaAs region, bottom of conduction
band confined on CrAs is not significantly changed by finite temperature
electronic correlations. Therefore the electrons and holes remain separated
in different regions of the real space.

\section{Summary}
Heterostructures made of layers of Heusler half-metals alternated with
semiconductors
suffer from the presence of interface states within 
the half-metallic gap which  strongly reduce the possibility of spin-polarized
current injection. From 
previous theoretical  studies it is known that 
the only case when 
half-metallicity 
is retained is the case of a NiMnSb(111)/CdS(111) 
interface formed between Sb and S atoms~\cite{wi.gr.01}. 
Alternative options are provided by  zinc-blende half-metals due to
their affinity with  the structure of zinc-blende semiconductors. In the present paper
we have performed combined electronic structure and many-body calculations 
for  short period (CrAs)$_n$/(GaAs)$_n$ heterostructures, with $n\le 3$ having
a  lattice parameter of $a=5.75 \AA$.
For $n=1$ we investigated the properties of a single-interface layer, and demonstrate that 
Cr-d$_{xy}$ and Cr-d$_{z^2-1}$ states that cross the Fermi level 
acquire a non-quasiparticle character in the presence of correlations.  
Our results demonstrate that heterostructures for $n=2,3$ 
feature
magnetizations  deviating by less than $2\%$ from the integer 
half-metallic values at temperatures close to  room temperature
$T=200, 300K$. The computed finite temperature polarizations deviate
significantly from the expected $100\%$. However,  we believe that these
values $P_{DMFT}(E_F)\approx 75\%$, that are far larger then values for 
other (semi-)Heuslers are still sufficient to 
make such a heterostructure interesting for future spintronic applications.  
In addition, we have investigated the 
character of the minority spin states around the Fermi level. While the top
of the valence band  mainly consists of As-p electrons localized on GaAs layers, 
the bottom of the conduction band is mainly determined by many-body induced NQP 
states having Cr-d character. 
This results, therefore, suggest that 
carriers remain separated among different layers
in these heterostructures.

We are grateful to D. Vollhardt for useful discussions. 
This work was supported by the Austrian science 
fund (FWF project P21289-N16)
and by the  cooperation project ``NAWI Graz''  (F-NW-515-GASS).

\bibliography{references_database}

\begin{thebibliography}{39}
\expandafter\ifx\csname natexlab\endcsname\relax\def\natexlab#1{#1}\fi
\expandafter\ifx\csname bibnamefont\endcsname\relax
  \def\bibnamefont#1{#1}\fi
\expandafter\ifx\csname bibfnamefont\endcsname\relax
  \def\bibfnamefont#1{#1}\fi
\expandafter\ifx\csname citenamefont\endcsname\relax
  \def\citenamefont#1{#1}\fi
\expandafter\ifx\csname url\endcsname\relax
  \def\url#1{\texttt{#1}}\fi
\expandafter\ifx\csname urlprefix\endcsname\relax\def\urlprefix{URL }\fi
\providecommand{\bibinfo}[2]{#2}
\providecommand{\eprint}[2][]{\url{#2}}

\bibitem[{\citenamefont{de~Groot et~al.}(1983)\citenamefont{de~Groot, Mueller,
  van Engen, and Buschow}}]{gr.mu.83}
\bibinfo{author}{\bibfnamefont{R.~A.} \bibnamefont{de~Groot}},
  \bibinfo{author}{\bibfnamefont{F.~M.} \bibnamefont{Mueller}},
  \bibinfo{author}{\bibfnamefont{P.~G.} \bibnamefont{van Engen}},
  \bibnamefont{and} \bibinfo{author}{\bibfnamefont{K.~H.~J.}
  \bibnamefont{Buschow}}, \bibinfo{journal}{Phys. Rev. Lett.}
  \textbf{\bibinfo{volume}{50}}, \bibinfo{pages}{2024} (\bibinfo{year}{1983}).

\bibitem[{\citenamefont{Katsnelson et~al.}(2008)\citenamefont{Katsnelson,
  Irkhin, Chioncel, Lichtenstein, and de~Groot}}]{ka.ir.08}
\bibinfo{author}{\bibfnamefont{M.~I.} \bibnamefont{Katsnelson}},
  \bibinfo{author}{\bibfnamefont{V.~Y.} \bibnamefont{Irkhin}},
  \bibinfo{author}{\bibfnamefont{L.}~\bibnamefont{Chioncel}},
  \bibinfo{author}{\bibfnamefont{A.~I.} \bibnamefont{Lichtenstein}},
  \bibnamefont{and} \bibinfo{author}{\bibfnamefont{R.~A.}
  \bibnamefont{de~Groot}}, \bibinfo{journal}{Reviews of Modern Physics}
  \textbf{\bibinfo{volume}{80}}, \bibinfo{pages}{315} (\bibinfo{year}{2008}).

\bibitem[{\citenamefont{Zhao and Zunger}(2005)}]{zh.zu.05}
\bibinfo{author}{\bibfnamefont{Y.-J.} \bibnamefont{Zhao}} \bibnamefont{and}
  \bibinfo{author}{\bibfnamefont{A.}~\bibnamefont{Zunger}},
  \bibinfo{journal}{Phys. Rev. B} \textbf{\bibinfo{volume}{71}},
  \bibinfo{pages}{132403} (\bibinfo{year}{2005}).

\bibitem[{\citenamefont{Akinaga et~al.}(2000)\citenamefont{Akinaga, Magano, and
  Shirai}}]{ak.ma.00}
\bibinfo{author}{\bibfnamefont{H.}~\bibnamefont{Akinaga}},
  \bibinfo{author}{\bibfnamefont{T.}~\bibnamefont{Magano}}, \bibnamefont{and}
  \bibinfo{author}{\bibfnamefont{M.}~\bibnamefont{Shirai}},
  \bibinfo{journal}{Jpn. J. Appl. Phys.} \textbf{\bibinfo{volume}{39}},
  \bibinfo{pages}{L1118} (\bibinfo{year}{2000}).

\bibitem[{\citenamefont{Mizuguchi et~al.}(2002)\citenamefont{Mizuguchi,
  Akinaga, Manago, Ono, Oshima, Shirai, Yuri, Lin, Hsieh, and Chen}}]{mi.ak.02}
\bibinfo{author}{\bibfnamefont{M.}~\bibnamefont{Mizuguchi}},
  \bibinfo{author}{\bibfnamefont{H.}~\bibnamefont{Akinaga}},
  \bibinfo{author}{\bibfnamefont{T.}~\bibnamefont{Manago}},
  \bibinfo{author}{\bibfnamefont{K.}~\bibnamefont{Ono}},
  \bibinfo{author}{\bibfnamefont{M.}~\bibnamefont{Oshima}},
  \bibinfo{author}{\bibfnamefont{M.}~\bibnamefont{Shirai}},
  \bibinfo{author}{\bibfnamefont{M.}~\bibnamefont{Yuri}},
  \bibinfo{author}{\bibfnamefont{H.~J.} \bibnamefont{Lin}},
  \bibinfo{author}{\bibfnamefont{H.~H.} \bibnamefont{Hsieh}}, \bibnamefont{and}
  \bibinfo{author}{\bibfnamefont{C.~T.} \bibnamefont{Chen}},
  \textbf{\bibinfo{volume}{91}}, \bibinfo{pages}{7917} (\bibinfo{year}{2002}).

\bibitem[{\citenamefont{Nagao et~al.}(2004)\citenamefont{Nagao, Shirai, and
  Miura}}]{na.sh.04}
\bibinfo{author}{\bibfnamefont{K.}~\bibnamefont{Nagao}},
  \bibinfo{author}{\bibfnamefont{M.}~\bibnamefont{Shirai}}, \bibnamefont{and}
  \bibinfo{author}{\bibfnamefont{Y.}~\bibnamefont{Miura}}, \bibinfo{journal}{J.
  Appl. Phys.} \textbf{\bibinfo{volume}{95}}, \bibinfo{pages}{6518}
  (\bibinfo{year}{2004}).

\bibitem[{\citenamefont{Akinaga and Mizuguchi}(2004)}]{ak.mi.04}
\bibinfo{author}{\bibfnamefont{H.}~\bibnamefont{Akinaga}} \bibnamefont{and}
  \bibinfo{author}{\bibfnamefont{M.}~\bibnamefont{Mizuguchi}},
  \bibinfo{journal}{Journal of Physics: Condensed Matter}
  \textbf{\bibinfo{volume}{16}}, \bibinfo{pages}{S5549} (\bibinfo{year}{2004}).

\bibitem[{\citenamefont{Gopalan et~al.}(1989)\citenamefont{Gopalan,
  Christensen, and Cardona}}]{go.ch.89}
\bibinfo{author}{\bibfnamefont{S.}~\bibnamefont{Gopalan}},
  \bibinfo{author}{\bibfnamefont{N.~E.} \bibnamefont{Christensen}},
  \bibnamefont{and} \bibinfo{author}{\bibfnamefont{M.}~\bibnamefont{Cardona}},
  \bibinfo{journal}{Phys. Rev. B} \textbf{\bibinfo{volume}{39}},
  \bibinfo{pages}{5165} (\bibinfo{year}{1989}).

\bibitem[{\citenamefont{Continenza et~al.}(2001)\citenamefont{Continenza,
  Picozzi, Geng, and Freeman}}]{co.pi.01}
\bibinfo{author}{\bibfnamefont{A.}~\bibnamefont{Continenza}},
  \bibinfo{author}{\bibfnamefont{S.}~\bibnamefont{Picozzi}},
  \bibinfo{author}{\bibfnamefont{W.~T.} \bibnamefont{Geng}}, \bibnamefont{and}
  \bibinfo{author}{\bibfnamefont{A.~J.} \bibnamefont{Freeman}},
  \bibinfo{journal}{Phys. Rev. B} \textbf{\bibinfo{volume}{64}},
  \bibinfo{pages}{085204} (\bibinfo{year}{2001}).

\bibitem[{\citenamefont{Zhao et~al.}(2002)\citenamefont{Zhao, Geng, Freeman,
  and Delley}}]{zh.ge.02}
\bibinfo{author}{\bibfnamefont{Y.-J.} \bibnamefont{Zhao}},
  \bibinfo{author}{\bibfnamefont{W.~T.} \bibnamefont{Geng}},
  \bibinfo{author}{\bibfnamefont{A.~J.} \bibnamefont{Freeman}},
  \bibnamefont{and} \bibinfo{author}{\bibfnamefont{B.}~\bibnamefont{Delley}},
  \bibinfo{journal}{Phys. Rev. B} \textbf{\bibinfo{volume}{65}},
  \bibinfo{pages}{113202} (\bibinfo{year}{2002}).

\bibitem[{\citenamefont{Xie et~al.}(2003)\citenamefont{Xie, Xu, Liu, and
  Pettifor}}]{xi.xu.03}
\bibinfo{author}{\bibfnamefont{W.-H.} \bibnamefont{Xie}},
  \bibinfo{author}{\bibfnamefont{Y.-Q.} \bibnamefont{Xu}},
  \bibinfo{author}{\bibfnamefont{B.-G.} \bibnamefont{Liu}}, \bibnamefont{and}
  \bibinfo{author}{\bibfnamefont{D.~G.} \bibnamefont{Pettifor}},
  \bibinfo{journal}{Phys. Rev. Lett.} \textbf{\bibinfo{volume}{91}},
  \bibinfo{pages}{037204} (\bibinfo{year}{2003}).

\bibitem[{\citenamefont{Galanakis and Mavropoulos}(2003)}]{ga.ma.03}
\bibinfo{author}{\bibfnamefont{I.}~\bibnamefont{Galanakis}} \bibnamefont{and}
  \bibinfo{author}{\bibfnamefont{P.}~\bibnamefont{Mavropoulos}},
  \bibinfo{journal}{Phys. Rev. B} \textbf{\bibinfo{volume}{67}},
  \bibinfo{pages}{104417} (\bibinfo{year}{2003}).

\bibitem[{\citenamefont{Shirai}(2003)}]{shir.03}
\bibinfo{author}{\bibfnamefont{M.}~\bibnamefont{Shirai}}, \bibinfo{journal}{J.
  Appl. Phys.} \textbf{\bibinfo{volume}{93}}, \bibinfo{pages}{6844}
  (\bibinfo{year}{2003}).

\bibitem[{\citenamefont{Chioncel et~al.}(2005)\citenamefont{Chioncel,
  Katsnelson, de~Wijs, de~Groot, and Lichtenstein}}]{ch.ka.05}
\bibinfo{author}{\bibfnamefont{L.}~\bibnamefont{Chioncel}},
  \bibinfo{author}{\bibfnamefont{M.~I.} \bibnamefont{Katsnelson}},
  \bibinfo{author}{\bibfnamefont{G.~A.} \bibnamefont{de~Wijs}},
  \bibinfo{author}{\bibfnamefont{R.~A.} \bibnamefont{de~Groot}},
  \bibnamefont{and} \bibinfo{author}{\bibfnamefont{A.~I.}
  \bibnamefont{Lichtenstein}}, \bibinfo{journal}{Phys. Rev. B}
  \textbf{\bibinfo{volume}{71}}, \bibinfo{pages}{085111}
  (\bibinfo{year}{2005}).

\bibitem[{\citenamefont{Edwards and Hertz}(1973)}]{ed.he.73}
\bibinfo{author}{\bibfnamefont{D.~M.} \bibnamefont{Edwards}} \bibnamefont{and}
  \bibinfo{author}{\bibfnamefont{J.~A.} \bibnamefont{Hertz}},
  \bibinfo{journal}{Journal of Physics F-Metal Physics}
  \textbf{\bibinfo{volume}{3}}, \bibinfo{pages}{2191} (\bibinfo{year}{1973}).

\bibitem[{\citenamefont{Irkhin and Katsnelson}(1990)}]{ir.ka.90}
\bibinfo{author}{\bibfnamefont{V.~Y.} \bibnamefont{Irkhin}} \bibnamefont{and}
  \bibinfo{author}{\bibfnamefont{M.~I.} \bibnamefont{Katsnelson}},
  \bibinfo{journal}{J. Phys.: Condens. Matter} \textbf{\bibinfo{volume}{2}},
  \bibinfo{pages}{7151} (\bibinfo{year}{1990}).

\bibitem[{\citenamefont{Chioncel
  et~al.}(2003{\natexlab{a}})\citenamefont{Chioncel, Katsnelson, de~Groot, and
  Lichtenstein}}]{ch.ka.03}
\bibinfo{author}{\bibfnamefont{L.}~\bibnamefont{Chioncel}},
  \bibinfo{author}{\bibfnamefont{M.~I.} \bibnamefont{Katsnelson}},
  \bibinfo{author}{\bibfnamefont{R.~A.} \bibnamefont{de~Groot}},
  \bibnamefont{and} \bibinfo{author}{\bibfnamefont{A.~I.}
  \bibnamefont{Lichtenstein}}, \bibinfo{journal}{Phys. Rev. B}
  \textbf{\bibinfo{volume}{68}}, \bibinfo{pages}{144425}
  (\bibinfo{year}{2003}{\natexlab{a}}).

\bibitem[{\citenamefont{Chioncel et~al.}(2007)\citenamefont{Chioncel, Allmaier,
  Yamasaki, Daghofer, Arrigoni, Katsnelson, and Lichtenstein}}]{ch.al.07}
\bibinfo{author}{\bibfnamefont{L.}~\bibnamefont{Chioncel}},
  \bibinfo{author}{\bibfnamefont{H.}~\bibnamefont{Allmaier}},
  \bibinfo{author}{\bibfnamefont{A.}~\bibnamefont{Yamasaki}},
  \bibinfo{author}{\bibfnamefont{M.}~\bibnamefont{Daghofer}},
  \bibinfo{author}{\bibfnamefont{E.}~\bibnamefont{Arrigoni}},
  \bibinfo{author}{\bibfnamefont{M.}~\bibnamefont{Katsnelson}},
  \bibnamefont{and}
  \bibinfo{author}{\bibfnamefont{A.}~\bibnamefont{Lichtenstein}},
  \bibinfo{journal}{Phys. Rev. B} \textbf{\bibinfo{volume}{75}},
  \bibinfo{pages}{140406} (\bibinfo{year}{2007}).

\bibitem[{\citenamefont{Chioncel et~al.}(2008)\citenamefont{Chioncel, Sakuraba,
  Arrigoni, Katsnelson, Oogane, Ando, Miyazaki, Burzo, and
  Lichtenstein}}]{ch.sa.08}
\bibinfo{author}{\bibfnamefont{L.}~\bibnamefont{Chioncel}},
  \bibinfo{author}{\bibfnamefont{Y.}~\bibnamefont{Sakuraba}},
  \bibinfo{author}{\bibfnamefont{E.}~\bibnamefont{Arrigoni}},
  \bibinfo{author}{\bibfnamefont{M.~I.} \bibnamefont{Katsnelson}},
  \bibinfo{author}{\bibfnamefont{M.}~\bibnamefont{Oogane}},
  \bibinfo{author}{\bibfnamefont{Y.}~\bibnamefont{Ando}},
  \bibinfo{author}{\bibfnamefont{T.}~\bibnamefont{Miyazaki}},
  \bibinfo{author}{\bibfnamefont{E.}~\bibnamefont{Burzo}}, \bibnamefont{and}
  \bibinfo{author}{\bibfnamefont{A.~I.} \bibnamefont{Lichtenstein}},
  \bibinfo{journal}{Phys. Rev. Lett.} \textbf{\bibinfo{volume}{100}},
  \bibinfo{pages}{086402} (\bibinfo{year}{2008}).

\bibitem[{\citenamefont{Shirai et~al.}(2003)\citenamefont{Shirai, Ikeuchi,
  Taguchi, and Akinaga}}]{sh.ik.03}
\bibinfo{author}{\bibfnamefont{M.}~\bibnamefont{Shirai}},
  \bibinfo{author}{\bibfnamefont{K.}~\bibnamefont{Ikeuchi}},
  \bibinfo{author}{\bibfnamefont{H.}~\bibnamefont{Taguchi}}, \bibnamefont{and}
  \bibinfo{author}{\bibfnamefont{H.}~\bibnamefont{Akinaga}},
  \bibinfo{journal}{J. Supercond.} \textbf{\bibinfo{volume}{16}},
  \bibinfo{pages}{27} (\bibinfo{year}{2003}).

\bibitem[{\citenamefont{Kino et~al.}(2003)\citenamefont{Kino, Aryasetiawan,
  Solovyev, Miyake, Ohno, and Terakura}}]{ki.ar.03}
\bibinfo{author}{\bibfnamefont{H.}~\bibnamefont{Kino}},
  \bibinfo{author}{\bibfnamefont{F.}~\bibnamefont{Aryasetiawan}},
  \bibinfo{author}{\bibfnamefont{I.}~\bibnamefont{Solovyev}},
  \bibinfo{author}{\bibfnamefont{T.}~\bibnamefont{Miyake}},
  \bibinfo{author}{\bibfnamefont{T.}~\bibnamefont{Ohno}}, \bibnamefont{and}
  \bibinfo{author}{\bibfnamefont{K.}~\bibnamefont{Terakura}},
  \bibinfo{journal}{Physica B} \textbf{\bibinfo{volume}{329-333}},
  \bibinfo{pages}{858} (\bibinfo{year}{2003}).

\bibitem[{\citenamefont{Geshi et~al.}(2006)\citenamefont{Geshi, Shirai,
  Kusakabe, and Suzuki}}]{ge.sh.06}
\bibinfo{author}{\bibfnamefont{M.}~\bibnamefont{Geshi}},
  \bibinfo{author}{\bibfnamefont{M.}~\bibnamefont{Shirai}},
  \bibinfo{author}{\bibfnamefont{K.}~\bibnamefont{Kusakabe}}, \bibnamefont{and}
  \bibinfo{author}{\bibfnamefont{N.}~\bibnamefont{Suzuki}},
  \bibinfo{journal}{Computational Materials Science}
  \textbf{\bibinfo{volume}{36}}, \bibinfo{pages}{143 } (\bibinfo{year}{2006}),
  \bibinfo{note}{proceedings of the Second Conference of the Asian Consortium
  for Computational Materials Science -- ACCMS-2}.

\bibitem[{\citenamefont{Chioncel
  et~al.}(2003{\natexlab{b}})\citenamefont{Chioncel, Vitos, Abrikosov, Kollar,
  Katsnelson, and Lichtenstein}}]{ch.vi.03}
\bibinfo{author}{\bibfnamefont{L.}~\bibnamefont{Chioncel}},
  \bibinfo{author}{\bibfnamefont{L.}~\bibnamefont{Vitos}},
  \bibinfo{author}{\bibfnamefont{I.~A.} \bibnamefont{Abrikosov}},
  \bibinfo{author}{\bibfnamefont{J.}~\bibnamefont{Kollar}},
  \bibinfo{author}{\bibfnamefont{M.~I.} \bibnamefont{Katsnelson}},
  \bibnamefont{and} \bibinfo{author}{\bibfnamefont{A.~I.}
  \bibnamefont{Lichtenstein}}, \bibinfo{journal}{Phys. Rev. B}
  \textbf{\bibinfo{volume}{67}}, \bibinfo{pages}{235106}
  (\bibinfo{year}{2003}{\natexlab{b}}).

\bibitem[{\citenamefont{Kotliar et~al.}(2006)\citenamefont{Kotliar, Savrasov,
  Haule, Oudovenko, Parcollet, and Marianetti}}]{ko.sa.06}
\bibinfo{author}{\bibfnamefont{G.}~\bibnamefont{Kotliar}},
  \bibinfo{author}{\bibfnamefont{S.~Y.} \bibnamefont{Savrasov}},
  \bibinfo{author}{\bibfnamefont{K.}~\bibnamefont{Haule}},
  \bibinfo{author}{\bibfnamefont{V.~S.} \bibnamefont{Oudovenko}},
  \bibinfo{author}{\bibfnamefont{O.}~\bibnamefont{Parcollet}},
  \bibnamefont{and} \bibinfo{author}{\bibfnamefont{C.~A.}
  \bibnamefont{Marianetti}}, \bibinfo{journal}{Rev. Mod. Phys.}
  \textbf{\bibinfo{volume}{78}}, \bibinfo{pages}{865} (\bibinfo{year}{2006}).

\bibitem[{\citenamefont{Kotliar and Vollhardt}(2004)}]{ko.vo.04}
\bibinfo{author}{\bibfnamefont{G.}~\bibnamefont{Kotliar}} \bibnamefont{and}
  \bibinfo{author}{\bibfnamefont{D.}~\bibnamefont{Vollhardt}},
  \bibinfo{journal}{Physics Today} \textbf{\bibinfo{volume}{57}},
  \bibinfo{pages}{53} (\bibinfo{year}{2004}).

\bibitem[{\citenamefont{Held}(2007)}]{held.07}
\bibinfo{author}{\bibfnamefont{K.}~\bibnamefont{Held}}, \bibinfo{journal}{Adv.
  Phys.} \textbf{\bibinfo{volume}{56}}, \bibinfo{pages}{829}
  (\bibinfo{year}{2007}).

\bibitem[{\citenamefont{Katsnelson and Lichtenstein}(2002)}]{ka.li.02}
\bibinfo{author}{\bibfnamefont{M.~I.} \bibnamefont{Katsnelson}}
  \bibnamefont{and} \bibinfo{author}{\bibfnamefont{A.~I.}
  \bibnamefont{Lichtenstein}}, \bibinfo{journal}{Eur. Phys. J. B}
  \textbf{\bibinfo{volume}{30}}, \bibinfo{pages}{9} (\bibinfo{year}{2002}).

\bibitem[{\citenamefont{Imada et~al.}(1998)\citenamefont{Imada, Fujimori, and
  Tokura}}]{im.fu.98}
\bibinfo{author}{\bibfnamefont{M.}~\bibnamefont{Imada}},
  \bibinfo{author}{\bibfnamefont{A.}~\bibnamefont{Fujimori}}, \bibnamefont{and}
  \bibinfo{author}{\bibfnamefont{Y.}~\bibnamefont{Tokura}},
  \bibinfo{journal}{Rev. Mod. Phys.} \textbf{\bibinfo{volume}{70}},
  \bibinfo{pages}{1039} (\bibinfo{year}{1998}).

\bibitem[{\citenamefont{Chioncel
  et~al.}(2006{\natexlab{a}})\citenamefont{Chioncel, Arrigoni, Katsnelson, and
  Lichtenstein}}]{ch.ar.06}
\bibinfo{author}{\bibfnamefont{L.}~\bibnamefont{Chioncel}},
  \bibinfo{author}{\bibfnamefont{E.}~\bibnamefont{Arrigoni}},
  \bibinfo{author}{\bibfnamefont{M.~I.} \bibnamefont{Katsnelson}},
  \bibnamefont{and} \bibinfo{author}{\bibfnamefont{A.~I.}
  \bibnamefont{Lichtenstein}}, \bibinfo{journal}{Phys. Rev. Lett.}
  \textbf{\bibinfo{volume}{96}}, \bibinfo{pages}{137203}
  (\bibinfo{year}{2006}{\natexlab{a}}).

\bibitem[{\citenamefont{Chioncel
  et~al.}(2006{\natexlab{b}})\citenamefont{Chioncel, Mavropoulos,
  Le{\v{z}ai\'c}, Bl{\"u}gel, Arrigoni, Katsnelson, and
  Lichtenstein}}]{ch.ma.06}
\bibinfo{author}{\bibfnamefont{L.}~\bibnamefont{Chioncel}},
  \bibinfo{author}{\bibfnamefont{P.}~\bibnamefont{Mavropoulos}},
  \bibinfo{author}{\bibfnamefont{M.}~\bibnamefont{Le{\v{z}ai\'c}}},
  \bibinfo{author}{\bibfnamefont{S.}~\bibnamefont{Bl{\"u}gel}},
  \bibinfo{author}{\bibfnamefont{E.}~\bibnamefont{Arrigoni}},
  \bibinfo{author}{\bibfnamefont{M.~I.} \bibnamefont{Katsnelson}},
  \bibnamefont{and} \bibinfo{author}{\bibfnamefont{A.~I.}
  \bibnamefont{Lichtenstein}}, \bibinfo{journal}{Phys. Rev. Lett.}
  \textbf{\bibinfo{volume}{96}}, \bibinfo{pages}{197203}
  (\bibinfo{year}{2006}{\natexlab{b}}).

\bibitem[{\citenamefont{Lichtenstein et~al.}(2001)\citenamefont{Lichtenstein,
  Katsnelson, and Kotliar}}]{li.ka.01}
\bibinfo{author}{\bibfnamefont{A.~I.} \bibnamefont{Lichtenstein}},
  \bibinfo{author}{\bibfnamefont{M.~I.} \bibnamefont{Katsnelson}},
  \bibnamefont{and} \bibinfo{author}{\bibfnamefont{G.}~\bibnamefont{Kotliar}},
  \bibinfo{journal}{Phys. Rev. Lett.} \textbf{\bibinfo{volume}{87}},
  \bibinfo{pages}{067205} (\bibinfo{year}{2001}).

\bibitem[{\citenamefont{Petukhov et~al.}(2003)\citenamefont{Petukhov, Mazin,
  Chioncel, and Lichtenstein}}]{pe.ma.03}
\bibinfo{author}{\bibfnamefont{A.~G.} \bibnamefont{Petukhov}},
  \bibinfo{author}{\bibfnamefont{I.~I.} \bibnamefont{Mazin}},
  \bibinfo{author}{\bibfnamefont{L.}~\bibnamefont{Chioncel}}, \bibnamefont{and}
  \bibinfo{author}{\bibfnamefont{A.~I.} \bibnamefont{Lichtenstein}},
  \bibinfo{journal}{Phys. Rev. B} \textbf{\bibinfo{volume}{67}},
  \bibinfo{pages}{153106} (\bibinfo{year}{2003}).

\bibitem[{\citenamefont{Jarrell}(1992)}]{jarr.92}
\bibinfo{author}{\bibfnamefont{M.}~\bibnamefont{Jarrell}},
  \bibinfo{journal}{Phys. Rev. Lett.} \textbf{\bibinfo{volume}{69}},
  \bibinfo{pages}{168} (\bibinfo{year}{1992}).

\bibitem[{\citenamefont{Di~Marco et~al.}(2009)\citenamefont{Di~Marco, Min\'ar,
  Chadov, Katsnelson, Ebert, and Lichtenstein}}]{ma.mi.09}
\bibinfo{author}{\bibfnamefont{I.}~\bibnamefont{Di~Marco}},
  \bibinfo{author}{\bibfnamefont{J.}~\bibnamefont{Min\'ar}},
  \bibinfo{author}{\bibfnamefont{S.}~\bibnamefont{Chadov}},
  \bibinfo{author}{\bibfnamefont{M.~I.} \bibnamefont{Katsnelson}},
  \bibinfo{author}{\bibfnamefont{H.}~\bibnamefont{Ebert}}, \bibnamefont{and}
  \bibinfo{author}{\bibfnamefont{A.~I.} \bibnamefont{Lichtenstein}},
  \bibinfo{journal}{Phys. Rev. B} \textbf{\bibinfo{volume}{79}},
  \bibinfo{pages}{115111} (\bibinfo{year}{2009}).

\bibitem[{\citenamefont{Leonov et~al.}(2008)\citenamefont{Leonov, Binggeli,
  Korotin, Anisimov, Stoji\ifmmode~\acute{c}\else \'{c}\fi{}, and
  Vollhardt}}]{le.bi.08}
\bibinfo{author}{\bibfnamefont{I.}~\bibnamefont{Leonov}},
  \bibinfo{author}{\bibfnamefont{N.}~\bibnamefont{Binggeli}},
  \bibinfo{author}{\bibfnamefont{D.}~\bibnamefont{Korotin}},
  \bibinfo{author}{\bibfnamefont{V.~I.} \bibnamefont{Anisimov}},
  \bibinfo{author}{\bibfnamefont{N.}~\bibnamefont{Stoji\ifmmode~\acute{c}\else
  \'{c}\fi{}}}, \bibnamefont{and}
  \bibinfo{author}{\bibfnamefont{D.}~\bibnamefont{Vollhardt}},
  \bibinfo{journal}{Phys. Rev. Lett.} \textbf{\bibinfo{volume}{101}},
  \bibinfo{pages}{096405} (\bibinfo{year}{2008}).

\bibitem[{\citenamefont{Leonov et~al.}(2010)\citenamefont{Leonov, Korotin,
  Binggeli, Anisimov, and Vollhardt}}]{le.ko.10}
\bibinfo{author}{\bibfnamefont{I.}~\bibnamefont{Leonov}},
  \bibinfo{author}{\bibfnamefont{D.}~\bibnamefont{Korotin}},
  \bibinfo{author}{\bibfnamefont{N.}~\bibnamefont{Binggeli}},
  \bibinfo{author}{\bibfnamefont{V.~I.} \bibnamefont{Anisimov}},
  \bibnamefont{and}
  \bibinfo{author}{\bibfnamefont{D.}~\bibnamefont{Vollhardt}},
  \bibinfo{journal}{Phys. Rev. B} \textbf{\bibinfo{volume}{81}},
  \bibinfo{pages}{075109} (\bibinfo{year}{2010}).

\bibitem[{\citenamefont{Held et~al.}(2001)\citenamefont{Held, McMahan, and
  Scalettar}}]{he.mc.01}
\bibinfo{author}{\bibfnamefont{K.}~\bibnamefont{Held}},
  \bibinfo{author}{\bibfnamefont{A.~K.} \bibnamefont{McMahan}},
  \bibnamefont{and} \bibinfo{author}{\bibfnamefont{R.~T.}
  \bibnamefont{Scalettar}}, \bibinfo{journal}{Phys. Rev. Lett.}
  \textbf{\bibinfo{volume}{87}}, \bibinfo{pages}{276404}
  (\bibinfo{year}{2001}).

\bibitem[{\citenamefont{Chioncel and Burzo}(2006)}]{ch.bu.06}
\bibinfo{author}{\bibfnamefont{L.}~\bibnamefont{Chioncel}} \bibnamefont{and}
  \bibinfo{author}{\bibfnamefont{E.}~\bibnamefont{Burzo}}, \bibinfo{journal}{J.
  Optoelectronics Adv. Mat.} \textbf{\bibinfo{volume}{8}},
  \bibinfo{pages}{1105} (\bibinfo{year}{2006}).

\bibitem[{\citenamefont{de~Wijs and de~Groot}(2001)}]{wi.gr.01}
\bibinfo{author}{\bibfnamefont{G.~A.} \bibnamefont{de~Wijs}} \bibnamefont{and}
  \bibinfo{author}{\bibfnamefont{R.~A.} \bibnamefont{de~Groot}},
  \bibinfo{journal}{Phys. Rev. B} \textbf{\bibinfo{volume}{64}},
  \bibinfo{pages}{020402} (\bibinfo{year}{2001}).

\end{thebibliography}

\end{document}